\documentclass[final,3p,twocolumn]{elsarticle}

\usepackage{amssymb}
\usepackage{amsmath}
\usepackage{xcolor}
\usepackage{siunitx}
\usepackage{booktabs}
\usepackage{multirow}

\usepackage{needspace}
\usepackage{xurl}

\journal{Imaging Neuroscience}

\begin{document}

\begin{frontmatter}



\title{General Microstructure Factor Analysis of Diffusion MRI in Gray-Matter Predicts Cognitive Scores}

\author[1,2]{Lucas Z. Brito}
\author[3]{Ryan P. Cabeen}
\author[4]{David H. Laidlaw}

\affiliation[1]{organization={Department of Physics, Harvard University},
                city={Cambridge}, state={MA}, postcode={02140}, country={USA}}
\affiliation[2]{organization={Department of Physics, Brown University},
                city={Providence}, state={RI}, postcode={02912}, country={USA}}
\affiliation[3]{organization={Laboratory of Neuro Imaging, USC Mark and Mary Stevens Neuroimaging and Informatics Institute, Keck School of Medicine of USC, University of Southern California},
                city={Los Angeles}, state={CA}, postcode={90033}, country={USA}}
\affiliation[4]{organization={Department of Computer Science, Brown University},
                city={Providence}, state={RI}, postcode={02912}, country={USA}}

\begin{abstract}
Diffusion magnetic resonance imaging (MRI) has revealed important insights into white matter microstructure, but its application to gray matter remains comparatively less explored. Here, we investigate whether global patterns of gray-matter microstructure can be captured through neurite orientation dispersion and density imaging (NODDI) and whether such patterns are predictive of cognitive performance. 
Using diffusion MRI and behavioral data from the Human Connectome Project Young Adult study, we derive region-averaged NODDI parameters and apply principal component analysis (PCA) to construct general gray-matter microstructure factors. We  find that the factor derived from isotropic volume fraction explained substantial inter-individual variability and was significantly correlated with specific cognitive scores collected from the NIH Toolbox. In particular, the isotropic volume fraction factor  is linked to reading and vocabulary performance and to cognitive fluidity.
Our findings demonstrate that PCA-based global indicators of gray-matter microstructure provide complementary markers of structure–function relationships, extending beyond region-specific analyses. Our results suggest that general microstructure factors may serve as  population-level exploratory biomarkers for studying cognition and cortical organization. 
\end{abstract}





\end{frontmatter}


\begin{figure*}
    \centering
    \includegraphics[width=1\linewidth]{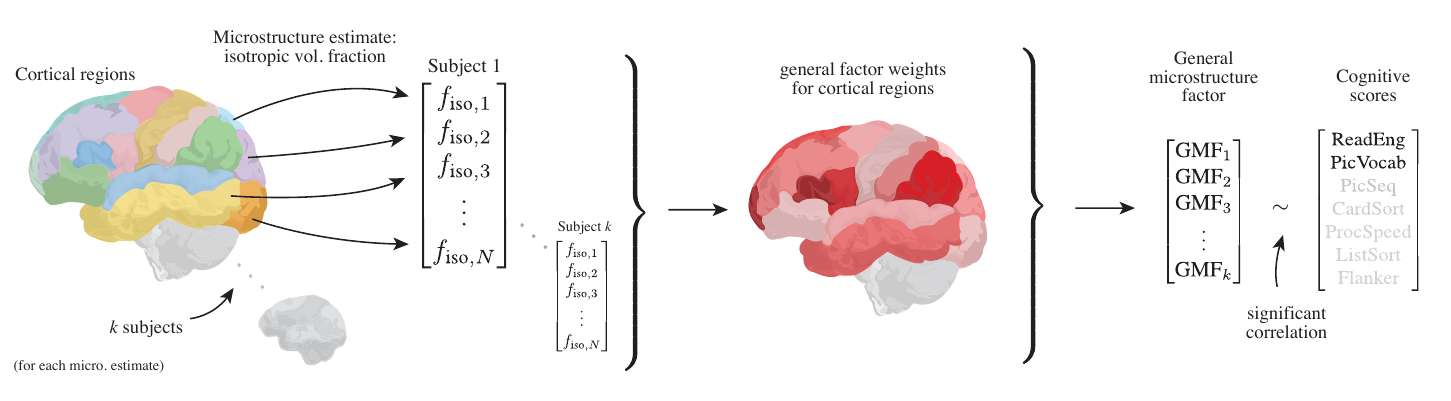}
    \caption{We study the relationship between global trends in NODDI gray-matter microstructure estimates and cognition using the procedure outlined  in} the figure. For each subject, we construct a vector with entries equal to per-region mean microstructure estimates. This produces, for each subject, a vector with one entry per cortical region. We then use  principal component analysis to compute a general microstructure factor (GMF) which assigns weights to each cortical region corresponding to that region's participation in global variation in  microstructure parameters. Each subject is thus assigned a GMF value. We then study trends between  the GMF and cognition using NIH Toolbox cognitive scores, and find a significant correlation between  the GMF and three cognitive scores.
    \label{fig:graphical-abstract}
\end{figure*}

\section{Introduction}\label{sec:introduction}

 The diffusion magnetic resonance imaging (MRI) study of gray-matter microstructure is an active field of research, comparatively less developed than diffusion MRI techniques aimed at white matter regions \cite{jones2010diffusion}. Diffusion imaging of gray matter is  more challenging for a number of reasons---for instance, the geometry of the cortical sheets makes voxels more sensitive to cerebrospinal fluid (CSF) partial volume effects, and the complex and comparatively isotropic tissue microstructure complicates detection of cellular architecture via  the diffusion signal \cite{jones2010diffusion, silva-rudberg}.
Nonetheless, success has been achieved in revealing gray-matter microstructure with diffusion scans, for instance at the structural level via sensitivity to laminar or regional microstructural heterogeneity \cite{Aggarwal2014-qy, Assaf2018-nm} or anisotropy in developing cortices \cite{Baratti1999-mo}. Thus, much inquiry has been directed at revealing correlations between microstructure and function via diffusion imaging, with a particular emphasis on cognition \cite{Radhakrishnan2022-qj, VENKATESH202012, radhakrishnan, Colgan2015-pz, Lynch2023-si}. 

One way to extract more specific information from diffusion signals is to fit the  signal to a multicompartment model. Although diffusion MRI acquisitions can provide, on their own, much anatomically relevant information to clinicians and researchers, often richer information is revealed by fitting acquisitions to biophysical models of tissue microstructure. In recent years, attention has been devoted to developing models  specifically capable of capturing the subtleties of gray matter \cite{PALOMBO2020116835}, but more generic models are also believed to reliably capture microstructural features. In this study, we employ one such model, neurite orientation dispersion and density imaging (NODDI), to study gray-matter microstructure \cite{ZHANG20121000}. 
NODDI is composed of an intracellular compartment that models diffusion in the region bounded by neurites, an extracellular compartment representing the vicinity of neurites, where  diffusion is restricted by, for example, glial cells or somas, and an isotropic compartment corresponding to CSF.
This is in contrast with more common imaging techniques such as diffusion tensor imaging (DTI), which is a single-compartment model consisting of a diffusivity tensor from which scalar measures such as fractional anisotropy (FA) are derived \cite{ODonnell2011-kk, BASSER1994247, Basser1994-ke}. In this sense,  DTI is best thought of as a signal representation as opposed to a multicompartment biophysical model \cite{Kamiya2020-tn}.

The use of NODDI in structural studies is promising, as it demonstrates greater specificity than  single-compartment DTI \cite{Grussu2017-tj, SCHILLING2018200}. For example, NODDI has been shown to capture  cognitive trends better than single-tensor approaches \cite{VENKATESH202012}. Consequently, it has become a common modality in studies of function–structure relationships centered on cognition and gray-matter microstructure. These include investigations of age-related changes in diffusivity  within hippocampal subregions \cite{VENKATESH202012}, associations between intracellular volume fraction, $f_{\text{icvf}}$, and cognitive deficits in patients with thalamic stroke \cite{Zhang2023-xv}, and links between neurite density and Alzheimer’s disease \cite{Parker2018-cp}. However, relatively few studies have examined global associations between NODDI-derived microstructural measures and cognitive performance. Certain tasks are likely to involve widespread networks, with population-level variation reflecting coordinated changes across multiple regions. Such analyses have been conducted in white matter tracts, where a general factor of tract integrity---computed using principal component analysis (PCA)---was found to correlate significantly with an analogous general factor of processing speed \cite{Penke7569}. In that case, approximately eight tracts contributed to a global indicator of white matter integrity. In this work, we build on that approach by computing a general factor across gray-matter regions and investigating its associations with cognitive performance ; see Fig. \ref{fig:graphical-abstract}. We call this measure the gray-matter general microstructure factor (GMF).  Gray matter introduces additional challenges and considerations: we employ NODDI as it is the simplest multi-compartment diffusion model capable of revealing relevant microstructural features, in contrast to the DTI-based approach used in the prior white matter study. Furthermore, the heightened sensitivity of gray matter to partial volume effects motivates us to residualize over region volume to control for these confounds. Our analysis is conducted on a large cohort of approximately one thousand participants from the Human Connectome Project Young Adult (HCP-YA) \cite{VANESSEN201362}.
Our paper is structured as follows: in Sec. \ref{sec:methods-data} we summarize  the HCP-YA acquisition protocol and preprocessing, as well as the cognitive behavioral scores used in our study. In Sec. \ref{sec:methods-analysis} we introduce the PCA-based analysis we perform on the NODDI maps. In Sec.~\ref{sec:results} we  present these results , and perform a series of posthoc tests to study the nature of these trends across  cortical regions.  In Sec.~\ref{sec:discussion} we discuss these findings. Lastly, we  summarize our findings in Sec.~\ref{sec:conclusion}.



\section{Methods}\label{sec:methods}
\subsection{DWI data}\label{sec:methods-data}
 Data were sourced from the Human Connectome Project Young Adult study (HCP-YA) 1200 subject release diffusion- and T1-weighted MRI acquisitions \cite{moeller-hcp}.  Data acquisition followed the HCP protocol described in \cite{VANESSEN201362, Glasser2013-zd}, which we review here. Diffusion-weighted MRI data were collected with the Connectome Skyra Siemens 3-Tesla scanner, using a 32-channel head coil. T1-weighted images were acquired with the 3D Magnetization Prepared - RApid Gradient Echo (MPRAGE) sequence with $\qty{0.7}{\milli\meter}$ isotropic resolution (field of view $\text{FOV}=\qty{224}{mm}$, in-plane matrix size $320$, and 256 slices in a single slab), with repetition time $T_R = \qty{2400}{\milli\second}$, echo time $T_E = \qty{2.14}{\milli\second}$, inversion time $T_I = \qty{1000}{\milli\second}$, $\text{flip angle} =8^\circ$, $\text{bandwidth} = \qty{210}{\hertz}$ per pixel, $\text{echo spacing} = \qty{7.6}{\milli\second}$, and phase encoding undersampling factor GeneRalized Autocalibrating Partial Parallel Acquisition $\text{GRAPPA}=2$.
Diffusion-weighted MRI images were acquired with a spin-echo EPI sequence with $\qty{1.25}{\milli\meter}$ isotropic resolution (FOV $\text{RO}\times\text{PE} = 210\times180$, matrix $\text{RO}\times\text{PE}= 168\times 144$ where RO is  the readout  direction and PE is  the phase-encoding  direction) with 111 slices of thickness $\qty{1.25}{\milli\meter}$ and a multiband factor of 3 with flip angles $68^\circ$ and $160^\circ$. Each phase-encoding direction utilized single-diffusion left-to-right and right-to-left (L/R,  R/L) encoding with $b=1000$, $2000$, and $\qty{3000}{\second\per\milli\meter\squared}$, sampled with 18 baseline scans and 270 diffusion-weighted scans with $T_E = \qty{89}{\milli\second}$ and $T_R = \qty{5520}{\milli\second}$. Each shell was acquired with 90 diffusion weighting directions, as well as an additional 6 $b=0$ acquisitions; these acquisitions are repeated twice for each encoding polarity.
Diffusion-weighted MRI images were preprocessed with the HCP pipeline detailed in \cite{Sotiropoulos2013-gd}. Relevant for our application is  the use of the L/R, R/L double phase encoding polarity to reduce motion, eddy-current, and susceptibility artifacts. These artifacts induce distortions which would otherwise prohibit accurate registration of diffusion-weighted and T1-weighted images, and therefore complicate alignment with cortical segmentation. Combining the two polarity-reversed scans allows researchers to estimate an off-resonance field which is then applied to produce a corrected image.

\subsection{Data processing}
The first stage of our analysis consisted of segmentation with Freesurfer and NODDI parameter fitting \cite{ZHANG20121000} partially carried out with the pipeline described in \cite{Cabeen2021-mk}. Specifically, we denoise diffusion-weighted MRI data with a non-local means filter, apply Freesurfer's eddy correction toolkit to remove eddy currents and perform motion correction, and fit NODDI parameters orientation dispersion index (ODI), isotropic volume fraction ($f_\text{iso}$) and fractional intracellular volume ($f_\text{icvf}$), and fiber orientation. We use a non-linear method accelerated via the spherical mean technique \cite{cabeen19}.
Since we specialize to gray-matter regions, we fit with a parallel diffusivity of \qty{1.1e-3}{\milli\meter\squared\per\second}, which  is optimized for gray-matter fitting. Microstructurally, ODI quantifies the variation of neurites' angular orientation in the voxel, $f_\text{iso}$ the fraction of the signal attributed to CSF, and $f_\text{icvf}$ the fraction attributed to the intracellular compartment, i.e., the space bounded by the neurite membranes. 
In this work, we utilize the Desikan-Killiany \cite{Desikan2006968} cortical atlas, which is a 34-region lateralized cortical parcellation.  Atlases were derived for each subject from T1-weighted images with Freesurfer v.~5.3.0.

\subsection{PCA analysis}\label{sec:methods-analysis}
\begin{figure*}
    \centering
    \includegraphics[width=1\linewidth]{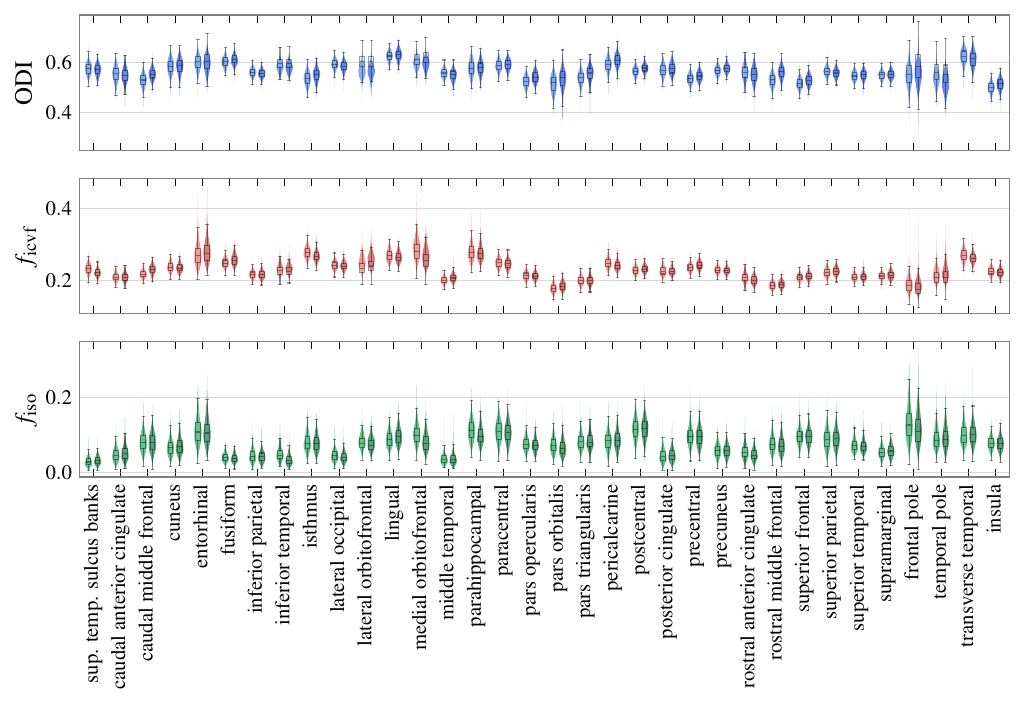}
    \caption{Distribution of mean per-region  microstructural values across subjects. Extremal values corresponding to misregistration or distortion artifacts have been filtered. This is an intermediate result; principal component analysis is performed on these distributions to produce the general factors with weights pictured in Fig.~\ref{fig:pca-weights}.  Mean distributions before data truncation are indicated with box plots.}
    \label{fig:params-per-reg}
\end{figure*}

Our aim is to extend the method of \cite{Penke7569} to gray-matter imaging by introducing a general gray-matter microstructure factor (GMF), constructed by  performing PCA on individual NODDI parameter estimates across gray-matter regions.
More specifically, each GMF is derived from region-averaged NODDI parameters $f_\text{iso}$, $f_\text{icvf}$ and ODI. PCA is performed along the dimension indexing cortical region, and the first principal component is selected as the general factor. See Fig. \ref{fig:params-per-reg} for the aggregated per-region NODDI estimates for all subjects. This generates, for each subject, three values corresponding to a weighted average of each NODDI parameter. We remind the reader that each subject's GMF is thus guaranteed to lie in the direction corresponding to maximal variance among subjects within the space of cortical regions. Furthermore, we note that by definition, all NODDI parameters are normalized to the unit interval $[0,1]$ and thus PCA sensitivity to scale variation is not a concern.
The role of this microstructure factor in driving cognitive effects is suggested by significant correlations with a subset of age-adjusted cognitive scores (obtained from HCP-YA data release). We utilize age-adjusted scores in our study  to control for age as a covariate.

We additionally truncate outlier voxels with extremal values of ODI and $f_\text{icvf}$ (below $0.03$ and above $0.95$). These values are anatomically unreasonable but can appear  as a result of misregistration, typically in inferior regions such as the entorhinal cortex and temporal pole , owing to distortion from the nasal cavity. Alternatively, spurious high values can appear in  individual voxels as artifacts from the NODDI fitting scheme, i.e., local minima in the optimization procedure.  Truncation is applied per-subject, and truncation bounds were chosen by directly inspecting the data, and lie outside  the bulk microstructural signal. For a typical subject this is at least $\approx1.6$ standard deviations away from  the mean NODDI parameter value. As we are only interested in  the mean value for each NODDI parameter, this truncation procedure eliminates outlier bias from our study.
    
One source of confounding when regressing against parameters derived from finite brain regions is the partial volume effect, or partial voluming. This is expected to be most pronounced for isotropic volume fraction $f_\text{iso}$ but may also confound other NODDI parameter estimates. Indeed, previous work has revealed that DTI parameters, for instance, are susceptible to partial voluming, and  researchers have suggested controlling for region volume \cite{VOS20111566}. 
In the present work we achieve this by residualizing NODDI parameters over region volume, i.e., fitting  an ordinary least-squares linear regression model with volume as predictor and microstructure parameter as response, and using the residuals of this model as input to our PCA pipeline. The regressions yielded average $r=-0.0513,-0.0949, 0.0634$ for ODI, $f_{\text{icvf}}$, and $f_{ \text{iso}}$ respectively .


\subsection{Behavioral measures}
Behavioral measures used in this study were sourced from the HCP-YA 1200 subject release behavioral dataset \cite{VANESSEN201362}. In this study, we specialize to NIH Toolbox tests under the ``Cognition'' category. These scores comprise oral reading (ReadEng), picture vocabulary (PicVocab), picture sequence (PicSeq), dimensional change card sorting (CardSort), pattern comparison processing speed (ProcSpeed), ``flanker'' (Flanker) and list sorting (ListSort) tests \cite{Akshoomoff, SeyedmehdiPayabvash}. We find significant correlations between  the general microstructure factor (GMF) and Flanker, ReadEng, and PicVocab.
The ``flanker'' task measures inhibitory control and attention abilities by asking participants to concentrate on one stimulus  while ignoring flanking stimuli;
 the oral reading test quantifies participants' abilities to accurately pronounce words and letters; the picture vocabulary test quantifies participants' vocabulary and general knowledge abilities by presenting subjects with a recording of a word and asking them to select one of four photos  that most closely match the word  heard. Each of these measures is normalized to national averages, such that a score of 100 denotes the national average, and one standard deviation is scaled to 15 points. Age-adjusted scores are derived by computing national averages for each age band; age bands are separated by year for ages 3--17, and into eight bands (18--29, 30--39, 40--49, 50--59, 60--69, 70--79, 80--85) for adults. We utilize age-adjusted scores to control for the correlation between age and NODDI diffusion measures, as  demonstrated for hippocampal regions in \cite{VENKATESH202012}.

\begin{table}[t]
\centering
\begin{tabular}{llll}
\toprule
 & ODI & $f_\mathrm{icvf}$ & $f_\mathrm{iso}$ \\
\midrule
Mean inter-region $r$ & 0.239 & 0.387 & 0.416 \\
Mean inter-region $r$ (lat.) & 0.249 & 0.390 & 0.421 \\
Explained variance & 0.232 & 0.355 & 0.401 \\
\bottomrule
\end{tabular}
\caption{Mean inter-region Pearson coefficients for each microstructural parameter. We include a lateralized mean Pearson coefficient where $r$ is not computed for  interhemispheric region  pairs, as analogous  contralateral regions are observed from Fig.~\ref{fig:params-per-reg} to be highly correlated. This may be interpreted as a quantification of the global correlation  in microstructural parameters throughout the cortex.}
\label{tab:pearson}
\end{table}

\Needspace{4\baselineskip}
\section{Results}\label{sec:results}

\begin{figure}
    \centering
    \includegraphics[width=1\linewidth]{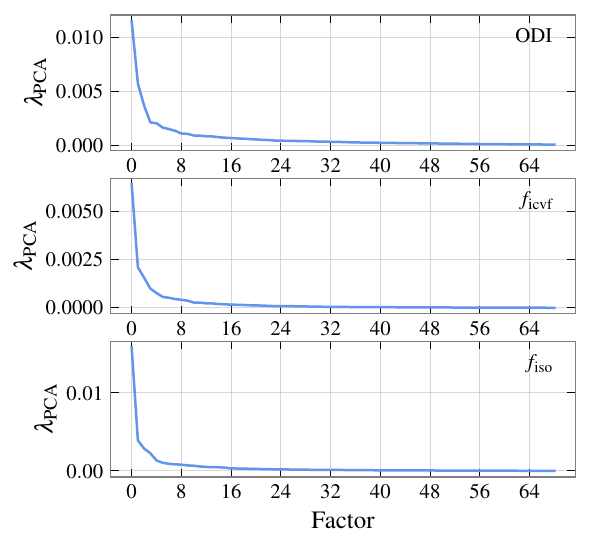}
    \caption{ Scree plots of the correlation matrix eigenvalues for the PCA of mean per-region microstructural parameters (see Fig.~\ref{fig:params-per-reg}).}
    \label{fig:scree}
\end{figure}

\begin{figure*}
    \centering
    \includegraphics[width=1\linewidth]{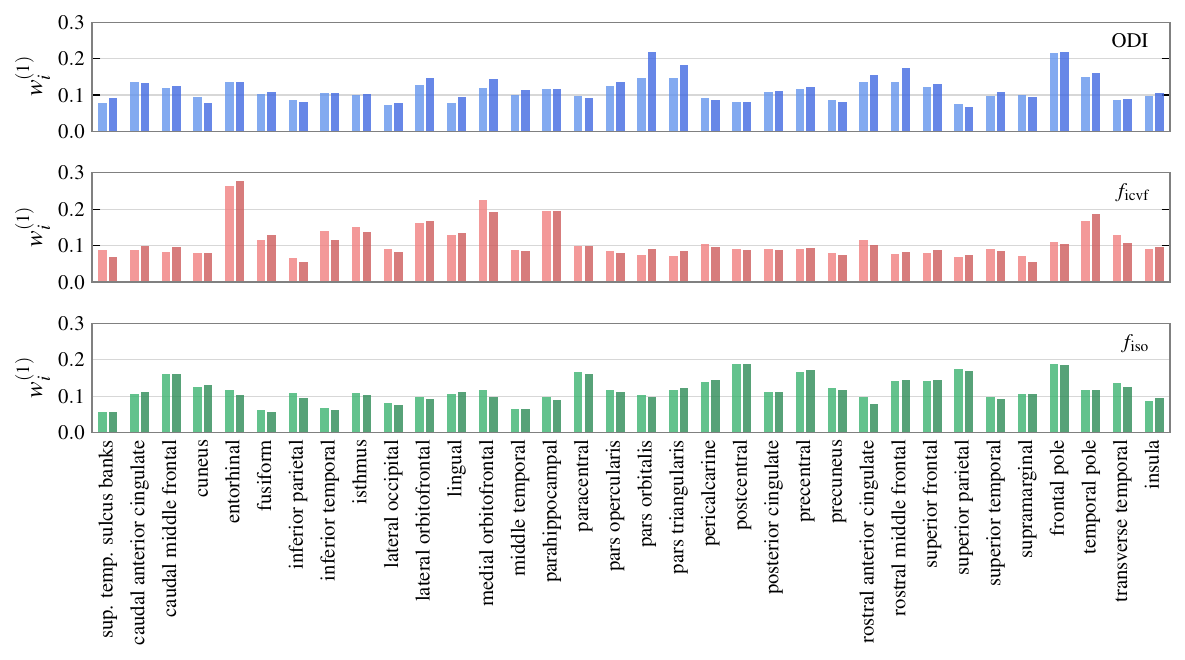}
    \caption{General factor weights corresponding to the first principal components of the distributions pictured in Fig.~\ref{fig:params-per-reg}. The height of each bar corresponds to the entry of  the eigenvector associated with that region, i.e., how much variation in microstructure for that region contributes to global variation.}
    \label{fig:pca-weights}
\end{figure*}

We perform principal component analysis (PCA) on each of the three scalar NODDI parameters, orientation dispersion index (ODI),  neurite compartment fraction ($f_\text{icvf}$), and CSF compartment fraction ($f_\text{iso}$).  Scree plots of the eigenvalues of the correlation matrix are shown in Fig. \ref{fig:scree}.  
No single ODI factor dominates the variance; indeed we find that the first factor explains only $22.6\%$ of the variance, suggesting a general ODI factor is less well defined than the other microstructure parameters. Specifically, $f_\text{icvf}$ and $f_\text{iso}$ have explained variances of $35.5\%$ and $40.1\%$ respectively, meaning that a single-factor solution is particularly strong for  the CSF compartment fraction. Factor weights are displayed in Fig. \ref{fig:pca-weights}. 
We additionally report average between-region Pearson coefficients $r$ in Table \ref{tab:pearson}. We include a lateralized computation ($r$ is not computed between regions in different hemispheres) to account for  the high correlation between analogous  contralateral regions.

We study the relationship between cognition and GMFs by performing linear regression with  the GMF as  the response variable and cognitive score as  the predictor. We standardize the data so that all measures are in units of the standard deviation, and study correlations between each GMF and picture vocabulary, oral reading, list and card sorting, flanker, picture sequence, and processing speed scores. We perform a multiple-comparisons correction on the $21$ $p$-values produced by these statistical tests. As the cognitive scores are positively correlated, we opt to use Benjamini-Hochberg false discovery rate (FDR) correction, which assumes the statistical tests are either independent or positively correlated \cite{Lindquist2015-hj, benjamini1995controlling}.

\begin{figure*}
    \centering
    \includegraphics[width=1\linewidth]{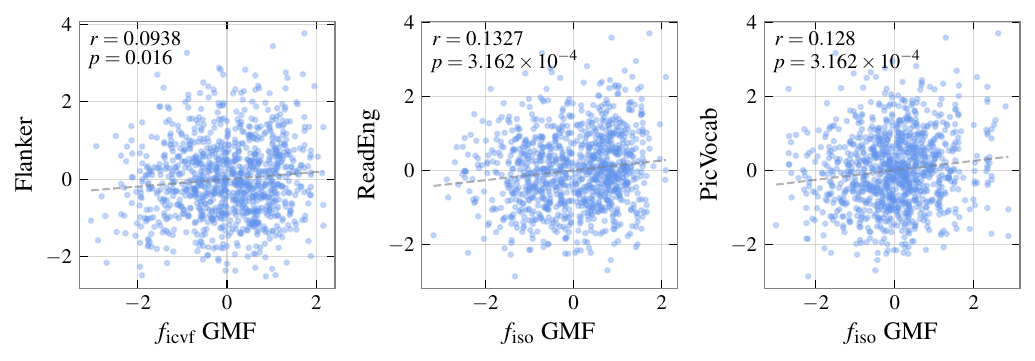}
    \caption{Lines of best fit for each of the significant trends observed. The dependent axis plots the general microstructure factor (GMF) of the corresponding NODDI parameter.  Reported $p$-values are multiple-comparisons-corrected.}
    \label{fig:fitted-lines}
\end{figure*}

We find statistically significant relationships between  the $f_\text{iso}$ GMF and ReadEng and PicVocab, and between the $f_\text{icvf}$ GMF and Flanker;
see Fig. \ref{fig:fitted-lines}. FDR-corrected $p$-values and $r$ coefficients are also displayed in Fig.~\ref{fig:fitted-lines}. Correlations between each of the GMFs and all other  cognitive scores are nonsignificant.
We study the structure of the significant trends observed by performing the following posthoc tests targeting trends across cortical regions.  We first examine associations between the significantly correlated cognitive scores and mean microstructure estimates for each cortical region. Then, we investigate the presence of trends that fail to be captured by the GMFs by, for each significant cognitive score and microstructure parameter, performing multiple regression over all regions with the inclusion of the GMF as a covariate.


We  study the detailed structure of the correlation between the cognitive scores and GMFs; in particular, we compute the correlations between microstructure estimate means for individual regions and the significant cognitive scores. We find that for
$f_\text{iso}$, 47 regions are significantly correlated with ReadEng ($1.901\times 10^{-6}<p<0.0478$) with trends comparable to GMF ($0.061<r<0.146$), 
and 51 regions are significantly correlated with PicVocab ($4.582\times 10^{-6}<p<0.0439$), likewise with trends comparable to the factor ($0.062<r<0.140$). We note PicVocab and ReadEng share 51 significant regions in common. 
On the other hand, $f_\text{icvf}$ has 39 regions significantly correlated with CardSort, with $1.741\times 10^{-6}<p < 0.0482 $ and $0.061<r<0.146$.
These results confirm that there is a significant trend in which many regions participate. To test how much of this trend is conveyed by the GMF , we control for the general factor and study the correlation between cognitive score and region estimate. We expect these correlations to reveal any trends not captured by the factor.
 After controlling for the GMF, $f_\text{icvf}$ correlates significantly with Flanker for 11 regions, with only 4 of those regions trending in the same direction as the factor.
Similarly, only 7 regions correlate $f_{ \text{iso}}$ significantly with ReadEng, and only 7 correlate significantly with PicVocab.

\section{Discussion}\label{sec:discussion}
We conclude that there is a significant structure-function relationship between global variation in cortical isotropic volume fraction $f_\text{iso}$ and performance in oral reading and picture vocabulary tests. 
A significant trend is also observed for the general $f_{\text{icvf}}$ factor and flanker test performance, although this is a weaker correlation , driven by associations with a smaller group of regions, indicating that the trend is less adequately captured by a global measure of tissue microstructure. This is reflected in the larger $p$-value for Flanker reported in Fig. \ref{fig:fitted-lines}.
We expect these results  to be robust to partial-voluming , thanks to residualization over region volume. 
Using generic F-test power analysis, we find the resulting $r$ values (Fig. \ref{fig:fitted-lines}) all have power greater than $0.80$ for $N = 1000$ and $\alpha = 0.05$ ($\text{power} = 0.87$, $0.99$, $0.99$ for Flanker, ReadEng and PicVocab respectively).  We make no claims of independence between  cognitive-structural associations, as our study is focused on pairwise trends between GMF and cognition. We leave open the possibility of collinearity, i.e., that a common cognitive mechanism underlies, e.g., both the ReadEng--$f_\text{iso}$ GMF and PicVocab--$f_\text{iso}$ GMF trends.  For instance, Flanker has correlation coefficients $0.16$ and $0.18$ for ReadEng and PicVocab respectively, and ReadEng and PicVocab have correlation coefficient $0.67$.

The observed trend relating $f_\text{iso}$ to cognition is expected as previous work has established, for example,  a relation between $f_\text{iso}$ and cognitive impairment \cite{Yu2024-go}, decreased $f_\text{iso}$ in language-processing regions for subjects with language impairments \cite{cabana2018effects}, and tau deposition in elderly cohorts \cite{schifani}.
 A potential microstructural mechanism is variation in perivascular space, which has been associated with cognition as well as isotropic diffusion compartments;  thus the $f_\text{iso}$ signal may be sensitive to perivascular space variation underlying cognitive fluidity and language processing, especially among young adults \cite{Choe2022-az, Sepehrband2019-yo, doi:10.1212/WNL.0000000000007124, Darnai2026}. We also leave open the possibility that this mechanism could be driven by a cell feature that is not captured by NODDI models; as discussed below, this motivates further investigation with diffusion models targeting gray matter specifically. We additionally observed that the $f_\text{icvf}$---Flanker association is more localized than the $f_\text{iso}$ associations, with the former displaying correlations across a smaller number of regions. Future work might employ, for instance, functional connectivity mapping analyses to understand the nature of this localization.

We close this section by listing some open questions raised by our analysis.
In this study we specialized to the NODDI microstructure model; however, several other models capable of capturing gray-matter microstructure are also available.  These include elementary signal representations such as diffusion tensor imaging (DTI), or, more relevant to the present case, more sophisticated models targeting gray matter, such as Soma and  Neurite Density Imaging (SANDI) \cite{PALOMBO2020116835}. One could carry out a multimodal study systematically comparing general factors derived from each model; this could shed light on the sensitivity of each model to global variations, as well as reveal whether such global variations in sophisticated models capture structure-function trends differently. We note that, for richer models such as SANDI, higher $b$-value acquisitions are required, and the  HCP-YA dataset used in this work would be insufficient  for this purpose.

We also mention that our analysis is constrained by the properties of PCA dimensionality reduction. While such linear approximations extract simple and interpretable trends, PCA assumes that the direction of maximal variation is a sound probe of anatomical signal and that inter-region trends are linearly correlated and orthogonal.  Consider the case where, for example, the direction of maximal variation is due to noise. It remains an open question whether better results might derive from more sophisticated dimensionality reduction techniques such as independent component analysis (ICA). A kernel principal component analysis (kPCA)  with a radial basis function (RBF) kernel yielded a first component with over $0.999$ correlation with the GMF for each of $f_\text{iso}$, $f_\text{icvf}$ , and ODI, and the same cognitive--structural correlation as our study. This suggests that a linear model is appropriate for this dataset. On the other hand, the first 4 ICA components , computed using FastICA, show no significant correlations with the GMF for typical initialization vectors. However, the ICA components exhibit significant variability across runs, as FastICA uses randomly-initialized weight vectors. Some runs yield components that are significantly correlated with different regions than those associated with the PCA-derived GMF, with $r$-values ranging from about $-0.20$ to $0.20$ and typical $p$-values of approximately $p=0.0003$--$0.0004$. This lack of robustness indicates that ICA does not yield stable or biologically meaningful components for this dataset.

Further, we encountered many subjects with small misregistrations that introduced artifacts to the dataset, typically originating in inferior regions. The method used in this work for omitting misregistered voxels was crude---we filter out voxels with microstructure estimates  outside a threshold range, motivated by anatomical constraints.  However, identifying the affected voxels by hand quickly becomes impractical in studies with a large number of subjects, such as ours. Thus, there is demand for a more refined and efficient technique for flagging misregistered regions. Additionally, although we applied state-of-the-art motion correction, there may be residual motion artifacts that remain \cite{Cieslak2022.07.21.500865}.

Lastly, we note that the technique explored in this study is amenable to several variations. While we generated general factors from trends across regions defined by the Desikan-Killiany atlas, one might compare  these against general factors derived from other segmentations.
Alternatively, one may sample microstructure parameters from regions of interest (ROIs) defined by surface maps, or eschew an ROI approach and sample voxels or surface map vertices directly. In the latter case, the dimensionality of the PCA problem can increase prohibitively; one may reduce the dimensionality by averaging microstructure parameters over subdivisions of a defined size, for example, spheres of a chosen radius. By inspecting the results of  the PCA  as the subdivision size is tuned, one may in principle study the dependence of  microstructural parameter trends on anatomical length scale. 

\section{Conclusion}\label{sec:conclusion}
In this work, we demonstrated that gray-matter general microstructure factors derived from NODDI parameters capture meaningful variation in cortical tissue organization and are significantly associated with specific cognitive domains. Using a large cohort from the Human Connectome Project, we showed that global $f_\text{iso}$ factors constructed via PCA explain a substantial portion of inter-individual variability. These factors were linked to performance in  oral reading and vocabulary tasks , with the $f_\text{iso}$ factor exhibiting broad cortical associations, and the $f_\text{icvf}$ factor showing more localized associations with flanker performance. Together, these findings highlight the utility of general microstructure factors as global indicators of cortical organization and suggest that they may provide a complementary framework for studying structure–function relationships beyond region-specific analyses.

\section*{Acknowledgments}
L. Z. B. acknowledges support from the Barry Goldwater Scholarship and Excellence in Education Foundation during part of this work. We thank Dr. Arthur Toga for providing the image computing resources through the Laboratory of Neuro Imaging Resource at USC (funded by NIH Grant No. P41EB015922).


 \bibliographystyle{elsarticle-num-names} 
 \bibliography{ref.bib}

\end{document}